# Origin of the higher-$T_\text{c}$ phase in the $K_x Fe_{2-y} Se_2$ system

Masashi TANAKA[1,‡], Yusuke YANAGISAWA[1,2], Saleem J. DENHOLME[1,*], Masaya FUJIOKA[1,**],

Shiro FUNAHASHI[1], Yoshitaka MATSUSHITA[1], Nobuo ISHIZAWA[3],

Takahide YAMAGUCHI[1], Hiroyuki TAKEYA[1], and Yoshihiko TAKANO[1,2]

[1]*National Institute for Materials Science, 1-2-1 Sengen, Tsukuba, Ibaraki 305-0047, Japan*

[2]*University of Tsukuba, 1-1-1 Tennodai, Tsukuba, Ibaraki 305-8577, Japan*

[3]*Advanced Ceramics Research Center, Nagoya Institute of Technology, 10-6-29 Asahigaoka,*

*Tajimi 507-0071, Japan*

[‡]Corresponding author: Masashi Tanaka,

E-mail: Tanaka.Masashi@nims.go.jp

Postal address: National Institute for Materials Science, 1-2-1 Sengen, Tsukuba, Ibaraki 305-0047, Japan

Tel.: (+81)29-851-3354 ext. 2976

[*]Present address: Department of Applied Physics, Tokyo University of Science, Shinjuku, Tokyo 162-8601, Japan

[**]Present address: Research Institute for Electronic Science, Hokkaido University, Sapporo, Hokkaido 001-0020, Japan

**Abstract**

Single crystals of $K_xFe_{2-y}Se_2$ are prepared by quenching at various temperatures. The crystals obtained at higher quenching temperatures have a surface morphology with mesh-like texture. They show a sharp superconducting transition at $T_c$ ~32 K with a large shielding volume fraction. On the other hand, the crystals prepared without quenching show an onset superconducting transition at ~44 K and a zero resistivity around ~33 K, and they possess island-like regions on the surface with a larger amount of Fe incorporation. *In-situ* high-temperature single crystal X-ray diffraction measurements tell us the Fe-vacancy ordered phase is generated at a temperature region around 270 °C via iron diffusion. The creation of this Fe-vacancy ordered phase may become a driving force of the growth of the higher $T_c$ phase. The superconductivity at ~44 K is attributed to a metallic phase with no Fe-vacancy.

## 1. Introduction

After the discovery of superconductivity in potassium intercalated FeSe [1], tremendous progress has been made in the study of related $A_x$Fe$_{2-y}$Se$_2$ ($A$ = K, Rb, Cs, Tl/Rb, Tl/K) systems [2-7] which exhibit relatively high superconducting (SC) transition temperatures $T_c$ ~30-46 K [8]. In the case of the potassium structures, a trace amount of a 44 K SC phase was observed which appeared to be separate from the 30 K phase in some samples [9,10], and even a ~48 K SC phase was reported under high pressure [11,12]. K-intercalation into FeSe via a liquid ammonia route has been reported to show SC at $T_c$ ~44 K [13]. These reports indicate that K$_x$Fe$_{2-y}$Se$_2$ is a potential superconductor with a maximum $T_c$ as high as 44-48 K. However, the question remains as to what causes the high $T_c$ phase with ~44 K to occur.

Many studies have shown that intrinsic phase separations occur in K$_x$Fe$_{2-y}$Se$_2$, leading to the coexistence of a SC phase with the ThCr$_2$Si$_2$-type (tetragonal, *I4/mmm*) structure and an Fe-vacancy ordered insulating K$_2$Fe$_4$Se$_5$ phase with √5 × √5 × 1 superstructure (tetragonal, *I4/m*), as shown in Figure 1 [14-26]. Recently, *in-situ* observation using scanning electron and transmission electron microscopies at elevated temperatures have shed some light on a potential growth mechanism for the SC phase in a single crystals [27,28]. However, the structural analysis of single crystals at high temperature has not been used to clarify the mechanism of superconductivity. In this study, we have evaluated the SC properties and surface morphology in K$_x$Fe$_{2-y}$Se$_2$ single crystals prepared under various conditions. The mechanism behind the production of the higher $T_c$ phase in the single crystal is discussed by means of *in-situ* high-temperature single crystal X-ray diffraction.

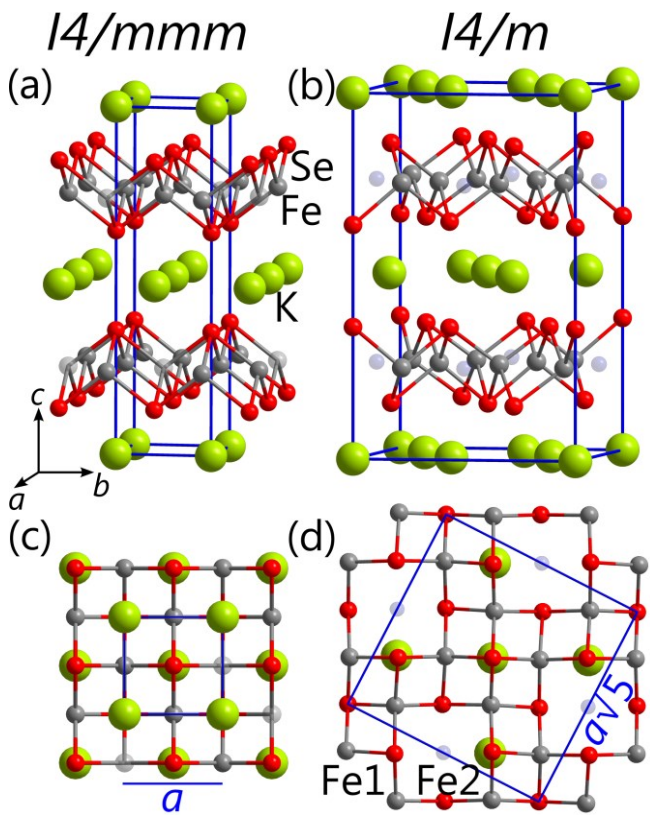

Figure 1. Schematic illustrations of unit cells for (a) the Fe-vacancy disordered $I4/mmm$, $K_xFe_{2-y}Se_2$, and (b) the Fe-vacancy ordered $I4/m$, $K_2Fe_4Se_5$ ($K_{0.8}Fe_{1.6}Se_2$), structures. (c), (d) The corresponding FeSe lattice viewing along $c$-axis direction. Transparent atoms indicate the Fe-vacancy in each structure.

## 2. Experimental

*2.1. Preparation of single crystals*

$K_xFe_{2-y}Se_2$ single crystals were prepared using a one-step method [29]. Powders of $K_2Se$, Fe, and Se grains were mixed with a nominal composition of $K_{0.8}Fe_2Se_2$ in an Ar atmosphere. The starting mixtures were placed in an alumina crucible, and were sealed in evacuated quartz tubes. The quartz tube was heated to 900 °C in 5 h, the temperature was then held for 12 h, followed by cooling to a specific quenching temperature or directly cooling to room temperature at a rate of 7 °C/h. All the sample preparation was performed in an Ar-filled glove box.

*2.2. Characterization*

The *in-situ* single crystal X-ray diffraction experiments (Mo K$\alpha$ radiation, $\lambda$ = 0.71073 Å) were carried out on several single crystals with dimensions of ~0.1 × ~0.2 × ~0.05 mm$^3$ at room temperature using a three-circle diffractometer with a CCD area detector (Smart Apex II, Bruker). Each crystal was sealed into an evacuated thin quartz capillary and the capillary was fixed with a ceramic adhesive. The sample temperature was controlled from room temperature up until 600 °C by a hot nitrogen gas stream [30]. The structure was solved by the Direct Method using SHELXS [31], and was refined with the program SHELXL [31] with the WinGX software package [32].

The temperature dependence of magnetization was measured using a SQUID magnetometer (MPMS, Quantum Design) down to 2 K under a field of 10 Oe, and the field was applied parallel to the *ab*-plane. The temperature dependence of electrical resistivity was measured down to 2.0 K, with a Physical Property Measurement System (PPMS, Quantum Design) using a standard four-probe method with constant current mode. The electrodes were attached in the *ab*-plane with silver paste. Back scattered electron (BSE) images and energy-dispersive X-ray (EDX) spectra were observed using a scanning electron microscope (JEOL, JSM-6010LA).

## 3. Results and discussion

*3.1 Superconducting properties*

Figure 2(a) shows the temperature dependence of the magnetic susceptibility of the crystal obtained by quenching at various temperatures. Crystals quenched at 700 °C show a sharp superconducting transition at $T_c$ ~32 K. The $T_c$ and shielding volume fraction tend to become lower and smaller, respectively, and also the transition width becomes broader, when decreasing the quenching temperature down to 300 °C. These tendencies are also observed in Ref. [33]. However, when the samples are cooled to room temperature without quenching (slow-cool), the onset temperature elevates to above 30 K, although the shielding fraction still remains low. This tendency was observed in all the samples quenched under 300 °C. Figure 2(b) shows the temperature dependence of resistivity of some selected crystals. The normal state conduction tends towards metallicity when increasing the quenching temperature. The slow-cooled sample has high resistivity and shows a broad hump around 180 K. In the enlargement of the transition temperature region (Figs. 2(c), (d)), the sample quenched at 700 °C shows a clear resistivity drop at ~31 K. However, the slow-cooled crystal shows a surprisingly high onset $T_c$ of 44 K and zero resistivity around 33 K, these values are clearly higher than those observed in the quenched crystals. This transition is suppressed by applying a magnetic field both in the direction along $H//ab$ and $H//c$. The Meissner signal was observed at ~43 K in the zero-field-cooling (ZFC) mode, and it depends on the applied magnetic field as shown in Figure 3. These phenomena both in the resistivity and the magnetization of the slow-cooled crystal suggest that the transition around 44 K is attributed to the superconducting transition.

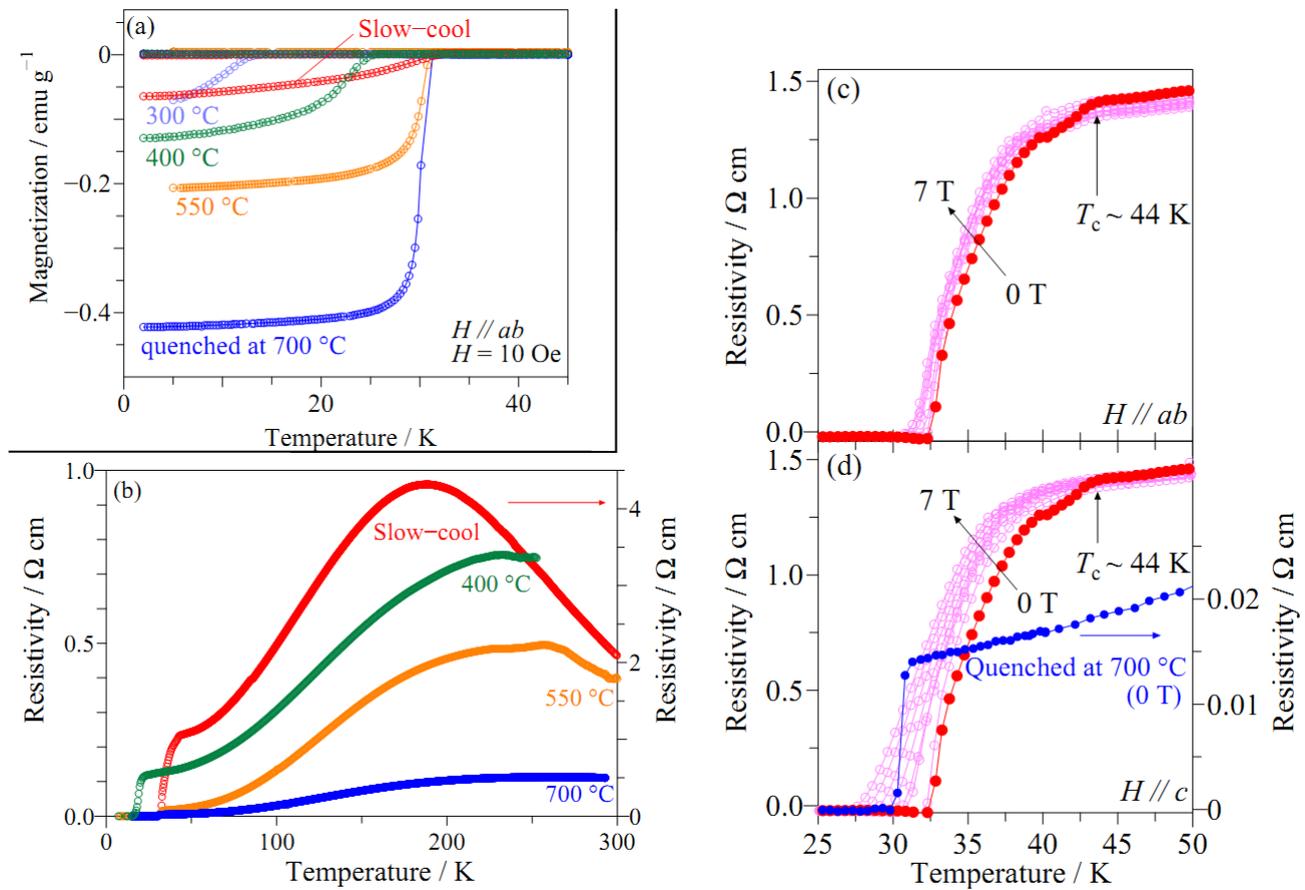

Figure 2. Temperature dependence of (a) magnetization, (b) resistivity of single crystal quenched at various temperatures. The enlarged scale of resistivity measurement under magnetic field parallel to *ab*-plane (c) and *c*-axis (d).

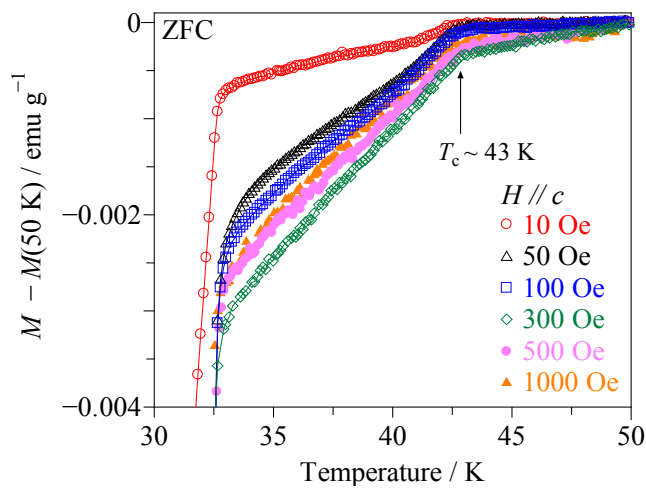

Figure 3. Temperature dependence of magnetization of slow-cooled crystal in enlarged scale under various magnetic fields parallel to *c*-axis. The magnetization values were subtracted by the value at 50 K.

*3.2 Surface morphology*

The quenching temperature also affects to the surface morphology of the single crystals as shown in Figure 4. The composition of the dark regions show stoichiometric values related to $K_{0.7}Fe_{1.61}Se_2$ for all crystals. The Fe to Se ratio close to 4:5 suggests that this phase corresponds to an insulating Fe-vacancy ordered phase, $K_2Fe_4Se_5$. On the other hand, the white contrasted region shows a different morphology between crystals quenched below and above 300 °C. Island-like regions in the slow-cooled crystal and mesh-like region for crystals quenched at 700 °C have a composition of $K_{0.40}Fe_{1.95}Se_2$ and $K_{0.63}Fe_{1.71}Se_2$, respectively. The island-like regions in the slow-cooled crystal contain more iron and less potassium than those of the mesh-like regions in the quenched crystals. The Fe content of the island-like region is almost 2.0, suggesting that the structure keeps its stoichiometry without any Fe-vacancies. A ratio between the white and dark regions in Figure 4 was estimated to be ~10-13% (< 300 °C) and ~30-35% (> 300 °C).

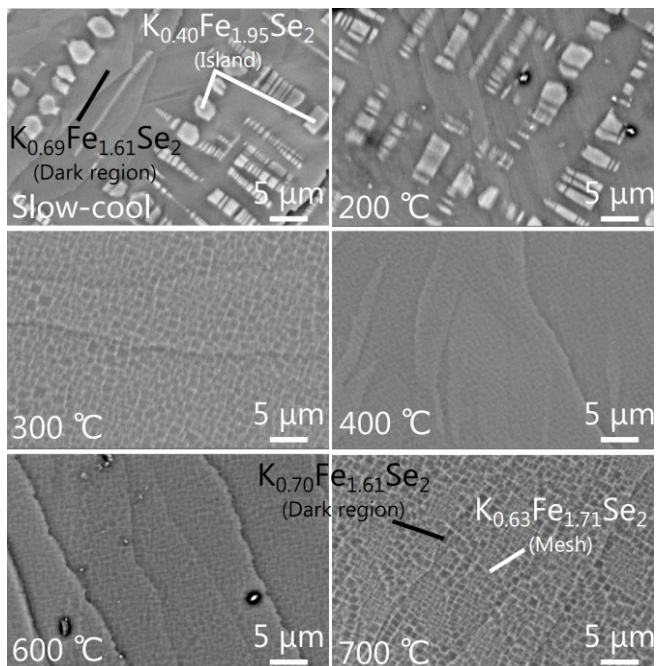

Figure 4. BSE images of the selected single crystals obtained by quenching at various temperatures.

*3.3 Single crystal X-ray analysis at various temperatures*

The single crystal X-ray diffraction experiments were firstly carried out at room temperature. One can see that the quenched crystal shows sharp diffraction spots in all directions suggesting high crystallinity as shown in Figure 5. However, the slow-cooled crystal shows highly diffused diffraction spots except for the *hk0* plane. All the reflections in *hk0* planes are labeled with Miller indices corresponding to the ThCr$_2$Si$_2$-type structure (space group *I4/mmm*) with lattice parameters $a = b$ ~3.9 Å. There are some Bragg peaks from the superstructure arising from the Fe-vacancy ordered phase with lattice parameters, $a = b$ ~8.7 Å and a lower symmetry of *I4/m*. The low crystallinity in the slow-cooled crystal may be attributed to a break in the long-range order along the *c*-axis caused by an unregulated stacking of layers during the growth of the single crystal.

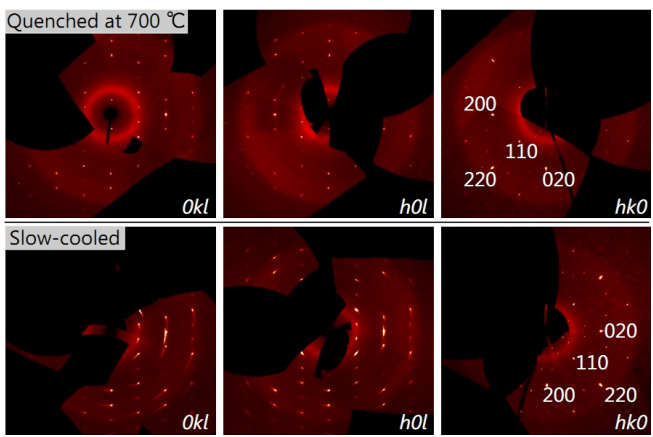

Figure 5. Reciprocal sections of single crystals at room temperature obtained by quenching at 700 °C (upper panels) and slow-cooled (lower panels) for each direction. All the indices are given on the basis of the *I4/mmm* structure cell.

Figure 6 shows the temperature dependence of the diffraction spots in the *hk0* section of the single crystal quenched at 700 °C. The superstructure diffraction can be seen below 250 °C; on the other hand, it completely disappears above 275 °C upon heating. The diffraction spots of the *I4/mmm* structure were confirmed to remain up to 600 °C while keeping its high crystallinity. When cooling down to 270 °C, the superstructure reflections start to resurface (marked by arrows), indicating that the Fe-vacancy order-disorder transition reversibly occurs. This reversible transition was also observed in the slow-cooled crystal in its *hk0* section. It is likely that quenching above 300 °C freezes the high temperature phase. The crystal structure was refined at each temperature and the details are shown in the supplementary information [34].

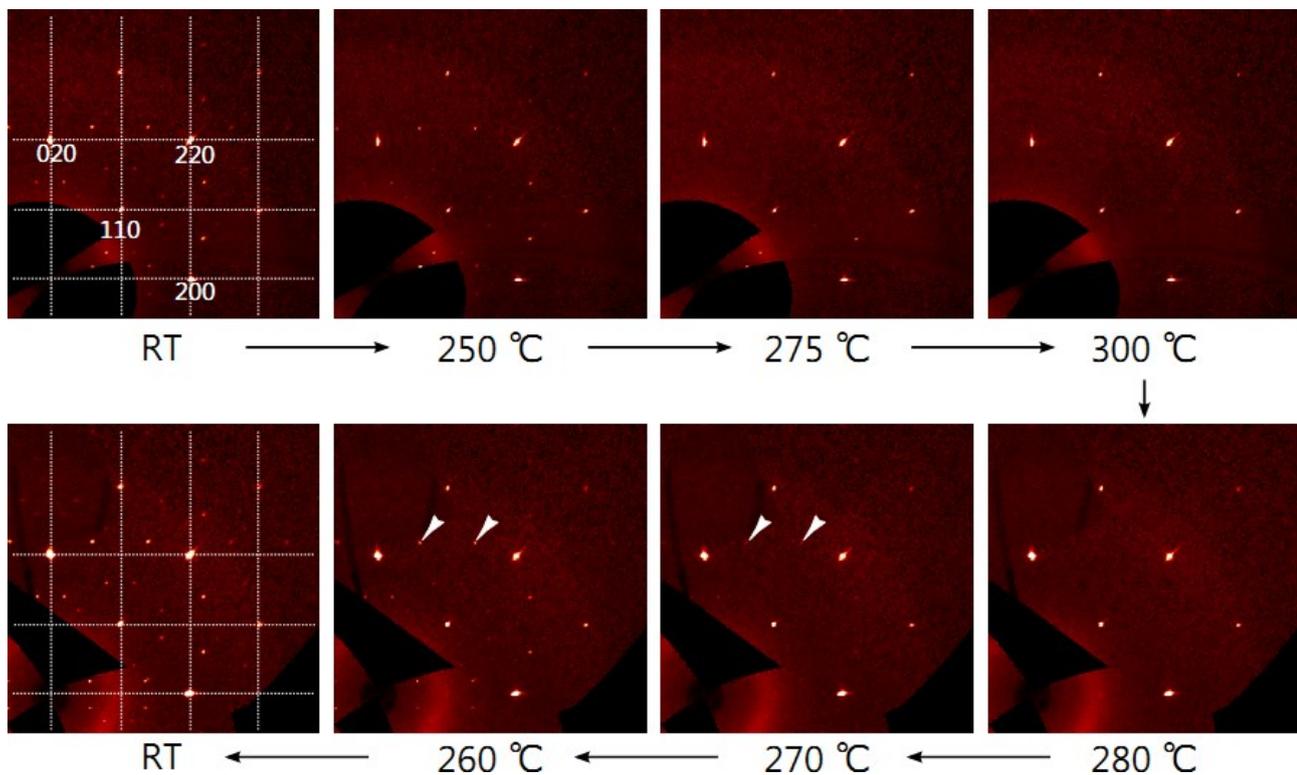

Figure 6. Reciprocal sections at elevated temperature for a single crystal quenched at 700 °C. The reflections are labeled with Miller indices corresponding to the *I4/mmm* structure. The white lines and arrows are a guide for eyes.

The reversible transition is also clearly seen in the temperature dependence of the refined lattice parameters $a$ and $c$, as shown in Figure 7(a). One can see a clear kink around ~270 °C in each parameter both when heating and cooling. The crystal tends to be compressed in the $ab$-plane and expanded along the $c$-axis by the growth of Fe-vacancy ordered phase below 270 °C. The elastic strain will accumulates along $c$-axis with an increase of the length of the SC strips [28].

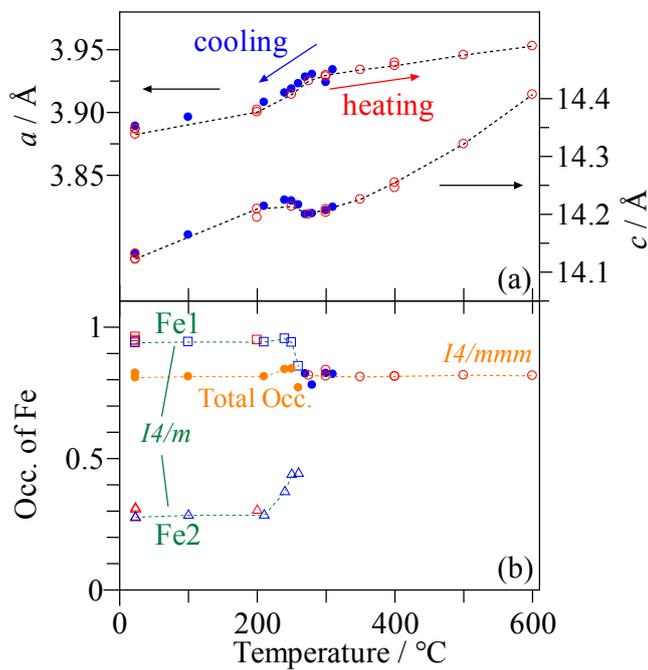

Figure 7. Temperature dependence of refined parameters: (a) the lattice parameters using a unit cell with the $I4/mmm$ structure, (b) the occupancy of Fe atoms in the crystals quenched at 700 °C. The red and blue colors indicate heating and cooling process, respectively. The unit cell in (b) was taken to the $I4/m$ and $I4/mmm$ structures below and above 260 °C, respectively. The total Occ. means averaged value of Fe sites. Fe1 and Fe2 correspond to the Fe sites shown in Figure 1(d).

The refined occupancy of the Fe-sites is plotted to Figure 7(b). In the *I4/m* structure, the Fe1 site is almost fully occupied, but the Fe2 site is occupied at about 30%. In an ideal insulating Fe-vacancy ordered phase, there should be full occupancy at the Fe1 site and full vacancy at Fe2 site [19,26]. The partial occupation of Fe2 site in this study indicates that ~30% of Fe atoms are randomly distributed in the crystal, suggesting that ~30% of the quenched crystal is composed of the Fe-vacancy disordered *I4/mmm* structure. We remind the reader that a ratio between mesh-like and dark regions in BSE images is ~30-35% of crystals quenched above 300 °C, which is in good agreement with the Fe2 occupancy.

Above 200 °C, the occupancy of Fe tends to merge into the averaged value, indicating that the Fe atoms can widely migrate in the given temperature range. The Fe2 site occupancy returns back to the same value after the heating cycle, a result which is not found in powder neutron diffraction studies [35]. This can be interpreted in the respect that the crystal size is large enough for lattice strain to prevent the escape of supersaturated Fe from the *I4/mmm* structure [19]. When the Fe-vacancy ordered phase appears, the excess Fe atoms will be expelled and diffuse into the *I4/mmm* structure while at the same time filling the Fe vacancy.

It has been reported that the mesh-like texture is a metallic phase having the electronic density of states at the Fermi level, and the dark region is a semiconducting/insulating phase [23]. In the quenched crystal, only a small amount of excess Fe can be incorporated into the *I4/mmm* structure at high temperatures because the cooling process is rapid enough to prevent Fe diffusion throughout the structure. So the mesh-like texture contains Fe vacancy in the crystal structure and shows superconductivity at $T_c$ ~32 K. On the other hand, the slow-cooled crystal was kept for a long time below 270 °C. The island-like region without Fe-vacancies will be formed and it may cause the occurrence of a higher $T_c$ superconductivity at 44

K. The migrated excess Fe is concentrated in the small isolated island-like regions, creating an inhomogeneous surface. The superconducting region is not large enough compare to its penetration depth. This also causes the suppression in the superconducting signal when applying magnetic field. This interpretation about a creation of the higher $T_c$ phase is not contradicted to an observation of superconducting transition at ~44 K in the K-Fe-Se system prepared by liquid ammonia technique [13]. A 'perfectly-Fe-incorporated' FeSe layers may have a potential to increase its superconducting transition up to ~44 K.

## 4. Conclusion

Single crystals of $K_xFe_{2-y}Se_2$ prepared by quenching at various temperatures showed different kind of superconductivity. One is a sharp superconducting transition at $T_c$ ~32 K in the mesh-like texture in the quenched crystal. The other is a zero resistivity at ~33 K and an onset superconducting transition at ~44 K occurred in the island-like regions in the slow-cooled crystals. They have a difference in the amount of Fe incorporation between mesh-like texture and the island-like regions, respectively. *In-situ* high-temperature single crystal X-ray diffraction measurements revealed that the Fe-vacancy ordered phase is generated at a temperature region around 270 °C with iron diffusion into the metallic *I4/mmm* structure. This may become a driving force of the growth of the higher $T_c$ phase. The superconductivity at ~44 K is attributed to the metallic phase with no Fe-vacancy. A further investigation for the direct analysis of the local structure and composition is required to prepare the single phase of superconductivity at 44 K.

**Acknowledgment**

This work was partially supported by Japan Science and Technology Agency through Strategic International Collaborative Research Program (SICORP-EU-Japan) and Advanced Low Carbon Technology R&D Program (ALCA) of the Japan Science and Technology Agency.

# Supplementary Information: Origin of the Higher-$T_c$ Phase in the K$_x$Fe$_{2-y}$Se$_2$ System

Masashi TANAKA[1], Yusuke YANAGISAWA[1,2], Saleem J. DENHOLME[1,*], Masaya FUJIOKA[1,**], Shiro FUNAHASHI[1], Yoshitaka MATSUSHITA[1], Nobuo ISHIZAWA[3], Takahide YAMAGUCHI[1], Hiroyuki TAKEYA[1], and Yoshihiko TAKANO[1,2]

[1]*National Institute for Materials Science, 1-2-1 Sengen, Tsukuba, Ibaraki 305-0047, Japan*
[2]*University of Tsukuba, 1-1-1 Tennodai, Tsukuba, Ibaraki 305-8577, Japan*
[3]*Advanced Ceramics Research Center, Nagoya Institute of Technology, 10-6-29 Asahigaoka, Tajimi 507-0071, Japan*

[*]*Present address: Department of Applied Physics, Tokyo University of Science, Shinjuku, Tokyo 162-8601, Japan*
[**]*Present address: Research Institute for Electronic Science, Hokkaido University, Sapporo, Hokkaido 001-0020, Japan*

Table S1. Summary of structural refinements for the quenched $K_xFe_{2-y}Se_2$ single crystals at various temperatures upon heating.

| Accession No. of Crystal | | #3 | #1 | #2 | #2 | #2 | #2 | #1 | #1 |
|---|---|---|---|---|---|---|---|---|---|
| Process | | heating | heating | heating | heating | heating | heating | heating | heating |
| Temperature (°C) | | 23 | 200 | 275 | 300 | 350 | 400 | 500 | 600 |
| Crystal System | | Tetragonal | Tetragonal | Tetragonal | Tetragonal | Tetragonal | Tetragonal | Tetragonal | Tetragonal |
| Space group | | $I4/m$ | $I4/m$ | $I/4mmm$ | $I/4mmm$ | $I/4mmm$ | $I/4mmm$ | $I/4mmm$ | $I/4mmm$ |
| $a$ (Å) | | 8.6918(6) | 8.7243(9) | 3.9251(4) | 3.9296(8) | 3.9339(4) | 3.9372(4) | 3.9457(2) | 3.9528(2) |
| $c$ (Å) | | 14.132(2) | 14.203(2) | 14.2002(14) | 14.209(3) | 14.2257(14) | 14.255(2) | 14.3215(12) | 14.4072(12) |
| $V$ (Å$^3$) | | 1067.6(2) | 1081.0(3) | 218.77(5) | 219.41(10) | 220.15(5) | 220.97(5) | 222.96(3) | 225.11(3) |
| | | | | | | | | | |
| Refinements | | | | | | | | | |
| $N_{meas}$ | | 1978 | 3059 | 440 | 429 | 439 | 459 | 668 | 674 |
| $2\theta_{max}$ (°) | | 60.01 | 60.05 | 59.66 | 59.61 | 59.95 | 59.81 | 60.35 | 60.18 |
| $R_{int}$ | | 0.0364 | 0.1076 | 0.0358 | 0.0357 | 0.0392 | 0.0437 | 0.0449 | 0.0391 |
| $N_{all}$ (independent) | | 767 | 829 | 122 | 122 | 123 | 123 | 131 | 131 |
| $N_{obs}$ ($>2\sigma(I)$) | | 359 | 271 | 103 | 103 | 102 | 97 | 90 | 89 |
| $R$ ($>2\sigma(I)$) | | 0.0516 | 0.0590 | 0.0372 | 0.0514 | 0.0479 | 0.0423 | 0.0425 | 0.0287 |
| $wR$ ($>2\sigma(I)$) | | 0.1300 | 0.1104 | 0.0835 | 0.1306 | 0.1238 | 0.0976 | 0.1043 | 0.0593 |
| $S_{all}$ | | 0.928 | 0.904 | 0.985 | 1.046 | 0.994 | 0.967 | 1.026 | 0.978 |
| $N_{par}$ | | 36 | 36 | 10 | 10 | 10 | 10 | 10 | 10 |
| $\Delta\rho_{max}$ (e/Å$^3$) | | 1.19 | 1.35 | 0.63 | 1.54 | 0.90 | 0.93 | 1.32 | 0.49 |
| $\Delta\rho_{min}$ (e/Å$^3$) | | -1.60 | -0.93 | -0.86 | -1.39 | -1.11 | -0.83 | -0.54 | -0.32 |
| | | | | | | | | | |
| Atom coordinates | | | | | | | | | |
| K 8h at $x, y$, 1/2 | Occ. | 0.71(2) | 0.74(2) | | | | | | |
| | $x$ | 0.0979(4) | 0.0979(7) | | | | | | |
| | $y$ | 0.6944(5) | 0.06944(8) | | | | | | |
| | $U_{eq}$ (Å$^2$) | 0.034(2) | 0.049(2) | | | | | | |
| K *2a* at 0,0,0 | Occ. | 0.72(3) | 0.76(3) | 0.78(2) | 0.80(3) | 0.79(3) | 0.76(3) | 0.72(3) | 0.68(2) |
| | $U_{eq}$ (Å$^2$) | 0.32(3) | 0.052(6) | 0.069(3) | 0.072(4) | 0.078(4) | 0.080(4) | 0.089(5) | 0.101(4) |
| Fe *16i* at *x, y, z* | Occ. | 0.953(9) | 0.951(7) | | | | | | |
| | x | 0.09331(13) | 0.0933(2) | | | | | | |
| | y | 0.19896(12) | 0.1998(2) | | | | | | |
| | z | 0.24761(13) | 0.2476(2) | | | | | | |
| | $U_{eq}$ (Å$^2$) | 0.0200(6) | 0.0297(7) | | | | | | |
| Fe *4d* at 1/2, 0, 1/4 | Occ. | 0.279(12) | 0.303(11) | 0.817(8) | 0.813(11) | 0.811(9) | 0.813(9) | 0.817(11) | 0.815(7) |
| | $U_{eq}$ (Å$^2$) | 0.018(3) | 0.032(5) | 0.0361(8) | 0.0374(11) | 0.0414(10) | 0.0431(10) | 0.0496(11) | 0.0585(8) |
| Se *16i* at *x,y,z* | x | 0.10804(8) | 0.10804(13) | | | | | | |
| | y | 0.70017(9) | 0.7006(2) | | | | | | |
| | z | 0.14501(8) | 0.14548(11) | | | | | | |
| | $U_{eq}$ (Å$^2$) | 0.0228(4) | 0.0305(4) | | | | | | |
| Se *4e* at 0,0,*z* | z | 0.3601(2) | 0.3593(3) | 0.35429(9) | 0.35421(12) | 0.35402(12) | 0.35374(11) | 0.35312(13) | 0.35268(9) |
| | $U_{eq}$ (Å$^2$) | 0.0217(6) | 0.0299(9) | 0.0386(5) | 0.0400(8) | 0.0438(7) | 0.0463(7) | 0.0512(7) | 0.0605(5) |

$N_{meas}$: number of measured reflections; $2\theta_{max}$: maximum $2\theta$ measured; $R_{int}$: merge $R$ value for equivalent reflections; $N_{all}$ (independent): number of independent reflections; $N_{obs}$ (>$2\sigma(I)$): number of independent reflections of significant intensity ($I > 2\sigma(I)$); $R$ (>$2\sigma(I)$): $R$ value for $N_{obs}$ reflections; $wR$ (>$2\sigma(I)$): weighted $R$ value for $N_{obs}$ reflections; $S_{all}$: the least-squares goodness-of-fit parameter for $N_{all}$ reflections; $N_{par}$: number of parameters refined; $\Delta\rho_{max}$ (e/Å$^3$): the largest value for the final difference electron density; $\Delta\rho_{min}$ (e/Å$^3$): the smallest value for the final difference electron density; Occ: occupation factor (not given if fully occupied).

Table S2. Summary of structural refinements for the quenched $K_xFe_{2-y}Se_2$ single crystals at various temperatures upon cooling.

| Accession No. of Crystal | | #3 | #2 | #3 | #3 | #3 | #3 | #3 | #3 | #3 | #3 |
|---|---|---|---|---|---|---|---|---|---|---|---|
| Process | | cooling | cooling | cooling | cooling | cooling | cooling | cooling | cooling | cooling | cooling |
| Temperature (°C) | | 310 | 300 | 280 | 270 | 260 | 250 | 240 | 210 | 100 | 23 |
| Crystal System | | Tetragonal | Tetragonal | Tetragonal | Tetragonal | Tetragonal | Tetragonal | Tetragonal | Tetragonal | Tetragonal | Tetragonal |
| Space group | | *I/4mmm* | *I/4mmm* | *I/4mmm* | *I/4mmm* | *I/4mmm* | *I4/m* | *I4/m* | *I4/m* | *I4/m* | *I4/m* |
| $a$ (Å) | | 3.9340(2) | 3.9240(6) | 3.9303(2) | 3.9282(2) | 3.9229(2) | 8.7627(4) | 8.7555(4) | 8.7394(5) | 8.7129(4) | 8.6961(5) |
| $c$ (Å) | | 14.2122(13) | 14.206(2) | 14.2010(13) | 14.1997(13) | 14.2163(13) | 14.2223(13) | 14.2230(14) | 14.213(2) | 14.162(2) | 14.130(2) |
| $V$ (Å$^3$) | | 219.95(3) | 218.74(8) | 219.37(3) | 219.11(3) | 218.78(3) | 1092.06(14) | 1090.32(15) | 1085.6(2) | 1075.1(2) | 1068.5(2) |
| | | | | | | | | | | | |
| Refinements | | | | | | | | | | | |
| $N_{meas}$ | | 525 | 448 | 519 | 524 | 531 | 2586 | 2328 | 1789 | 1770 | 1567 |
| $2\theta_{max}$ (°) | | 60.01 | 60.04 | 60.06 | 60.07 | 59.99 | 60.06 | 60.07 | 60.01 | 60.02 | 60.01 |
| $R_{int}$ | | 0.0189 | 0.0444 | 0.0165 | 0.0203 | 0.0235 | 0.0400 | 0.0390 | 0.0371 | 0.0321 | 0.0307 |
| $N_{all}$ (independent) | | 122 | 123 | 123 | 122 | 122 | 818 | 804 | 738 | 736 | 703 |
| $N_{obs}$ (>2$\sigma(I)$) | | 105 | 106 | 110 | 109 | 109 | 282 | 293 | 274 | 320 | 309 |
| $R$ (>2$\sigma(I)$) | | 0.0323 | 0.0473 | 0.0656 | 0.0281 | 0.0242 | 0.0500 | 0.0508 | 0.0595 | 0.0559 | 0.0494 |
| $wR$ (>2$\sigma(I)$) | | 0.0760 | 0.1226 | 0.1982 | 0.0712 | 0.0488 | 0.1131 | 0.1148 | 0.1302 | 0.1255 | 0.1009 |
| $S_{all}$ | | 1.182 | 1.059 | 1.264 | 1.215 | 1.085 | 0.919 | 0.900 | 0.957 | 0.988 | 0.906 |
| $N_{par}$ | | 10 | 10 | 10 | 10 | 10 | 36 | 36 | 36 | 36 | 36 |
| $\Delta\rho_{max}$ (e/Å$^3$) | | 0.64 | 1.18 | 2.09 | 0.66 | 0.55 | 1.45 | 1.42 | 1.39 | 1.25 | 1.45 |
| $\Delta\rho_{min}$ (e/Å$^3$) | | -0.53 | -1.12 | -0.91 | -1.10 | -0.71 | -0.91 | -1.03 | -1.11 | -1.23 | -1.41 |
| | | | | | | | | | | | |
| Atom coordinates | | | | | | | | | | | |
| K *8h* at $x, y, 1/2$ | Occ. | | | | | | 0.69(2) | 0.71(2) | 0.74(3) | 0.72(2) | 0.75(2) |
| | $x$ | | | | | | 0.0973(5) | 0.0979(5) | 0.0969(6) | 0.0986(5) | 0.0989(4) |
| | $y$ | | | | | | 0.6955(5) | 0.6951(5) | 0.6948(6) | 0.6938(6) | 0.6939(5) |
| | $U_{eq}$ (Å$^2$) | | | | | | 0.058(2) | 0.057(2) | 0.057(3) | 0.041(2) | 0.038(2) |
| K $2a$ at 0,0,0 | Occ. | 0.73(2) | 0.75(3) | 0.88(5) | 0.71(2) | 0.722(13) | 0.74(2) | 0.77(2) | 0.76(3) | 0.72(3) | 0.77(2) |
| | $U_{eq}$ (Å$^2$) | 0.065(3) | 0.069(4) | 0.074(5) | 0.060(2) | 0.060(2) | 0.057(4) | 0.063(5) | 0.049(5) | 0.038(4) | 0.034(3) |
| Fe $16i$ at $x, y, z$ | Occ. | | | | | | 0.942(6) | 0.956(7) | 0.943(10) | 0.944(9) | 0.941(8) |
| | $x$ | | | | | | 0.09512(13) | 0.09471(13) | 0.0934(2) | 0.0935(2) | 0.09373(14) |
| | $y$ | | | | | | 0.20029(12) | 0.20005(12) | 0.1999(2) | 0.19927(14) | 0.19908(14) |
| | $z$ | | | | | | 0.24823(16) | 0.24802(15) | 0.2474(2) | 0.2476(2) | 0.24742(15) |
| | $U_{eq}$ | | | | | | 0.0356(6) | 0.0354(6) | 0.0321(8) | 0.0241(6) | 0.0205(5) |
| Fe $4d$ at 1/2, 0, 1/4 | Occ. | 0.821(8) | 0.824(9) | 0.780(14) | 0.822(7) | 0.821(5) | 0.439(8) | 0.373(9) | 0.284(13) | 0.284(11) | 0.277(10) |
| | $U_{eq}$ (Å$^2$) | 0.0394(8) | 0.0383(10) | 0.0318(14) | 0.0363(7) | 0.0349(5) | 0.028(2) | 0.027(3) | 0.026(4) | 0.016(3) | 0.015(3) |
| Se $16i$ at $x,y,z$ | $x$ | | | | | | 0.10596(8) | 0.10668(9) | 0.10764(11) | 0.10783(10) | 0.10769(9) |
| | $y$ | | | | | | 0.70062(9) | 0.70085(9) | 0.70056(12) | 0.70037(10) | 0.7004(1) |
| | $z$ | | | | | | 0.14581(8) | 0.14565(8) | 0.14561(11) | 0.14508(9) | 0.14505(8) |
| | $U_{eq}$ (Å$^2$) | | | | | | 0.0357(4) | 0.0358(4) | 0.0347(5) | 0.0271(4) | 0.0234(4) |
| Se $4e$ at 0,0,$z$ | $z$ | 0.35415(9) | 0.35405(11) | 0.35417(15) | 0.35434(8) | 0.35461(6) | 0.3567(2) | 0.3572(2) | 0.3588(2) | 0.3598(2) | 0.3601(2) |
| | $U_{eq}$ (Å$^2$) | 0.0416(5) | 0.0407(7) | 0.0368(10) | 0.0383(4) | 0.0372(3) | 0.0342(6) | 0.0333(7) | 0.0314(8) | 0.0260(7) | 0.0219(6) |

$N_{meas}$: number of measured reflections; $2\theta_{max}$: maximum $2\theta$ measured; $R_{int}$: merge $R$ value for equivalent reflections; $N_{all}$ (independent): number of independent reflections; $N_{obs}$ (>2$\sigma$(I)): number of independent reflections of significant intensity ($I > 2\sigma(I)$); $R$ (>2$\sigma$(I)): $R$ value for $N_{obs}$ reflections; $wR$ (>2$\sigma$(I)): weighted $R$ value for $N_{obs}$ reflections; $S_{all}$: the least-squares goodness-of-fit parameter for $N_{all}$ reflections; $N_{par}$: number of parameters refined; $\Delta\rho_{max}$ (e/Å$^3$): the largest value for the final difference electron density; $\Delta\rho_{min}$ (e/Å$^3$): the smallest value for the final difference electron density; Occ: occupation factor (not given if fully occupied).